\documentclass[10pt]{article}

\usepackage[a4paper, inner=1.5cm, top=1.5cm, outer=1.5cm, bottom=1.5cm, includefoot]{geometry}
\usepackage{amsmath}
\usepackage{bm}
\usepackage[max2]{authblk}
\usepackage{amssymb}
\usepackage{mathrsfs}
\usepackage{graphicx}
\usepackage{enumitem}
\usepackage{tikz-network} 
\usepackage[ruled,lined]{algorithm2e}
\usepackage[style=ieee,natbib=true,url=false,isbn=false,mincitenames=1,maxcitenames=2]{biblatex}
\usepackage{amsthm}
\usepackage{appendix}
\usepackage{subdepth}
\usepackage{float}
\usepackage{hyperref}
\usepackage{mathtools}
\usepackage{multirow}
\usepackage{array}

\DeclareUnicodeCharacter{03B3}{$\gamma$}
\DeclareUnicodeCharacter{2212}{--}
\DeclareUnicodeCharacter{03B1}{$\alpha$}

\newcommand{\sty}[1]{\boldsymbol{#1}}
\newcommand{\styy}[1]{\mathbb{#1}}
\newcommand{\ubar}[1]{\mkern 0.5mu\underline{\mkern-0.5mu#1\mkern-0.5mu}\mkern 0.5mu}
\newcommand{\uubar}[1]{\ubar{\ubar{#1}}}

\let\epsilon\varepsilon
\let\mtheta\theta
\let\theta\vartheta

\let\rho\varrho

\let\phi\varphi
\let\Gamma\varGamma
\let\Delta\varDelta
\let\Theta\varTheta
\let\Lambda\varLambda
\let\Xi\varXi
\let\Pi\varPi
\let\Sigma\varSigma
\let\Upsilon\varUpsilon
\let\Phi\varPhi
\let\Psi\varPsi

\let\Omega\varOmega

\newcommand{\bmb}{\sty{b}}
\newcommand{\bmc}{\sty{c}}

\newcommand{\bmf}{\sty{f}}

\newcommand{\bmk}{\sty{k}}

\newcommand{\bmr}{\sty{r}}

\newcommand{\bmu}{\sty{u}}
\newcommand{\bmv}{\sty{v}}

\newcommand{\bmx}{\sty{x}}

\newcommand{\bmsigma}{\sty{\sigma}}

\newcommand{\bmZero}{\sty{0}}

\newcommand{\bbR}{\styy{R}}

\newcommand{\clE}{\mathcal{E}}

\newcommand{\clN}{\mathcal{N}}

\newcommand{\sT}{\mathsf{T}} 

\newcommand{\scC}{\mathscr{C}}

\newcommand{\scL}{\mathscr{L}}

\newcommand{\scN}{\mathscr{N}}

\newcommand{\scT}{\mathscr{T}}

\newcommand{\rme}{\mathrm{e}}
\newcommand{\rmf}{\mathrm{f}}

\newcommand{\rmx}{\mathrm{x}}
\newcommand{\rmy}{\mathrm{y}}
\newcommand{\rmz}{\mathrm{z}}

\newcommand{\uc}{\ubar{c}}

\newcommand{\umtheta}{\ubar{\mtheta}}

\newcommand{\uuA}{\uubar{A}}

\newcolumntype{x}[1]{>{\centering\arraybackslash\hspace{0pt}}p{#1}}

\title{Automated ab initio-accurate atomistic simulations of dissociated dislocations}
\author[1]{L. Mismetti\thanks{laura.mismetti@epfl.ch}}
\author[2]{M. Hodapp\thanks{corresponding author}\thanks{maxludwig.hodapp@mcl.at}}
\affil[1]{École polytechnique fédérale de Lausanne (EPFL), Lausanne (CH)}
\affil[2]{Materials Center Leoben Forschung GmbH (MCL), Leoben (AT)}

\DeclareUnicodeCharacter{03B3}{$\gamma$}
\DeclareUnicodeCharacter{2212}{--}
\DeclareUnicodeCharacter{03B1}{$\alpha$}

\begin{document}

\maketitle

\begin{abstract}
    In (M Hodapp and A Shapeev 2020 Mach. Learn.: Sci. Technol. 1 045005), we have proposed an algorithm that fully automatically trains machine-learning interatomic potentials (MLIPs) during large-scale simulations, and successfully applied it to simulate screw dislocation motion in body-centered cubic tungsten.
    The algorithm identifies local subregions of the large-scale simulation region where the potential extrapolates, and then constructs periodic configurations of 100--200 atoms out of these non-periodic subregions that can be efficiently computed with plane-wave Density Functional Theory (DFT) codes.

    In this work, we extend this algorithm to dissociated dislocations with arbitrary character angles and apply it to partial dislocations in face-centered cubic aluminum.
    Given the excellent agreement with available DFT reference results, we argue that our algorithm has the potential to become a universal way of simulating dissociated dislocations in face-centered cubic and possibly also other materials, such as hexagonal closed-packed magnesium, and their alloys.
    Moreover, it can be used to construct reliable training sets for MLIPs to be used in large-scale simulations of curved dislocations.
\end{abstract}

\section{Introduction}

Dislocations in metals are atomistic defects and, therefore, simulation methods that operate on the atomic scale are necessary in order to predict their behavior with the best possible accuracy.
Simulating a dislocation requires configurations of at least 100--200 atoms and the currently most accurate method that allows to simulate such configurations is considered to be Kohn-Sham density functional theory (in the following also just DFT).
The, to date, arguably most popular implementations of DFT use plane-wave basis sets, as implemented in codes like VASP \cite{kresse_efficient_1996};
for instance, the Materials Project database (\href{https://materialsproject.org/}{materialsproject.org}) is almost entirely based on plane-wave DFT calculations.
With such configurations, dislocations in body-centered cubic (bcc) metals can be simulated as dipoles, subject to periodic boundary conditions, because dislocations in bcc metals have a compact core structure (see, e.g., \cite{bigger_atomic_1992,romaner_effect_2010}).
For dislocations in face-centered cubic (fcc) metals, the situation is different: fcc dislocations dissociate into partial dislocations, with a splitting distance of several times the magnitude of the Burgers vector $b$ (cf., \cite{hirth_theory_1982}).
For example, for nickel, the partial splitting is around 8--10$b$ (see, e.g., \cite{szajewski_influence_2018}), which requires supercells of at least 2\,000--3\,000 thousand atoms to accommodate a dislocation dipole.
Such a large number of atoms presently still appears to be prohibitively expensive using plane-wave DFT, because plane-wave DFT scales cubically with the number of particles.
However, to be predictive, simulation methods must be able to simulate the partial splitting very accurately as it influences other mechanisms, such as dislocation cross-slip.
To that end, e.g., \citet{woodward_prediction_2008,tan_dislocation_2019,bianchini_enabling_2019} used quantum mechanics/molecular mechanics (QM/MM) methods that resolve only the dislocation core fully atomistically and use elasticity in the far-field.
Other works by \citet{iyer_electronicstructure_2015,das_electronic_2017} have used more efficient orbital-free (OF-DFT) methods that allow to simulate much larger cells with more than 5\,000 atoms.

However, none of the previous methods presently appears to be tractable to solve more complex problems that require millions of atoms, e.g., dislocations in high-entropy alloys, or long, curved dislocations.
QM/MM methods using plane-wave DFT as the QM model would still require too many atoms to resolve the dislocation core; moreover, they require additional buffer and vacuum regions to be included in the DFT supercell to ensure accurate force fields in the vicinity of the dislocation core.
OF-DFT methods have, to date, not been able to accurately predict the energetics of dislocations in alloys, in general (cf., e.g., \cite{dan_firstprinciples_2022}).
Newer state-of-the-art implementations of Kohn-Sham DFT with finite-element basis sets offer a high scalability compared to plane-wave DFT, allowing for configurations with 10\,000--100\,000 atoms, while achieving the same accuracy (cf., \citep{motamarri_dftfe_2020,das_largescale_2023}).
However, they still require supercomputers to simulate such a large number of atoms.

In this work, we approach the problem of simulating fcc dislocations with machine-learning potentials (MLIPs) \cite{behler_generalized_2007,bartok_gaussian_2010,thompson_spectral_2015,shapeev_moment_2016,schutt_schnet_2017,smith_ani1_2017,pun_physically_2019,drautz_atomic_2019,batzner_equivariant_2022,takamoto_teanet_2022}, parametrized using plane-wave DFT calculations.
Contrary to empirical interatomic potentials, which are in general not quantitatively accurate, MLIPs have a flexible functional form that allows to systemtically approximate local DFT energies down to the usual noise in numerical DFT codes \cite{shapeev_moment_2016}.
In practice, it is observed that electronic effects in many metals decay sufficiently fast so that properties that are relevant for computational metallurgy, such as dislocation core structures, core energies, etc., are not quantitatively impacted when replacing DFT with an interatomic interaction model.
Next-generation MLIPs even include nonlocal interactions or charges (e.g., \citep{gilmer_neural_2017,behler_machine_2021,takamoto_teanet_2022}) that allows to account for more complex quantum-mechanical effects if, e.g., the assumption of local interactions does not hold.
All of this makes MLIPs a very promising candidate for predictive large-scale simulations for dislocations in alloys.

One of the success stories of MLIPs is their ability to generalize to large systems:
they can be trained on sets of configurations of a few hundred atoms, yet they still make excellent predictions when simulating much larger configurations of tens of thousands of atoms (see, e.g., \citep{stricker_prismatic_2020,poul_systematic_2023,meziere_accelerating_2023}, and for some recent mathematical analysis of this observation \citep{ortner_framework_2023,wang_theoretical_2023}).
On the other hand, one of the major challenges in constructing a good MLIP is the construction of such a training set.
The vast majority of works are focusing on developing so-called general-purpose potentials that must be trained on huge datasets containing various defects (vacancies, stacking faults, grain boundaries, etc.) \cite{dragoni_achieving_2018,pun_development_2020,goryaeva_efficient_2021,marchand_machine_2022,deng_largescale_2023}.
Such potentials are predictive as long as their range of application is not too far from the training data, but not necessarily DFT-accurate in regions where it has not been trained on (cf., e.g., \cite{fellinger_geometries_2018}).

However, for alloy design, it is more tractable to use constitutive models for a mechanical property of interest (strength, ductility, hardness, etc.) that depend on a set of descriptors, e.g., the interaction energy of dislocation with a solute, to screen for the best possible alloy.
Therefore, it is also highly desirable to develop efficient \emph{special-purpose} potentials in order to keep the training set small-sized and the potential sufficiently reliable for that specific descriptor.
However, it is difficult to construct such a training set because it is, as outlined above, usually not feasible to simulate extended defects, such as dislocations, with plane-wave DFT alone.

To that end, we have developed an active learning algorithm for large-scale simulations of dislocations using moment tensor potentials (MTPs) \cite{hodapp_operando_2020}.
Our algorithm uses D-optimality to measure the per-atom uncertainty---the extrapolation grade---of the MTP in the simulation region, and extracts those subregions in which the extrapolation grade exceeds some threshold.
Then, our algorithm completes these local (in general non-periodic) subregions to \emph{periodic configurations} that are suitable to be computed with plane-wave DFT.
We have applied this method to simulate screw dislocation motion in bcc tungsten and obtained excellent agreement with reference DFT results.

In this work, we extend this algorithm to general dissociated dislocations with arbitrary character angles.
The new challenge, compared to our previous work on compact dislocations, is to extract \emph{several} subregions from the dislocation core and complete them to periodic configurations of 100--200 atoms.
To that end, we develop a method that extracts two regions, one around each partial dislocation core, and completes them to two periodic configurations with a net partial Burgers vector of zero in which no artificial neighborhoods occur at the periodic cell boundaries that could possibly degrade the accuracy of the MTP.
Using our algorithm, we train several MTPs on dislocations in fcc aluminum and compare their core structures and core energies to available DFT results from the literature.

\section{Methodology}

Before explaining our methodology, we fix some notation.
Let $\{\bmr_i\}_{i=1,\ldots,N} = \{\bmr_i\}$ be an arbitrary (possibly non-periodic) configuration of $N$ atoms, which represents the system at a specific step of the atomistic simulation.
The neighborhood of the $i$-th atom is the set $\scN = \{ \bmr_{ij} \}$ of all relative distances $\bmr_{ij} = \bmr_j - \bmr_i$ between $\bmr_i$ and atoms $\bmr_j$ within a few lattice spacings.
We assume that the total energy $\Pi$ of $\{\bmr_i\}$ is partitioned into per-atom contributions $\clE = \clE(\scN)$ such that
\begin{equation}\label{eq:total_energy}
    \Pi = \Pi(\{\bmr_i\}) = \sum_{\scN \in \{\bmr_i\}} \clE(\scN).
\end{equation}

\subsection{Moment Tensor Potentials}

We model the per-atom energies with Moment Tensor Potentials (MTPs) \cite{shapeev_moment_2016,gubaev_accelerating_2019}, which are based on a linear combination of basis function $B_\alpha$
\begin{equation}\label{eq:per-atom_energy}
    \clE(\scN) = \sum_\alpha \mtheta_\alpha B_\alpha(\scN),
\end{equation}
where the $\mtheta_\alpha$'s are free parameters.
The basis functions are built from scalar contractions of the moment tensors
\begin{equation}\label{eq:mom-tens-desc}
    M_{\mu, \nu}(\scN)
    =
    \sum_{\boldsymbol{r}_{i j} \in \scN} \sum_n c_{\mu n} f_n\left(\left|\boldsymbol{r}_{i j}\right|\right)
    \big( \underbrace{\boldsymbol{r}_{i j} \otimes \cdots \otimes \boldsymbol{r}_{i j}}_{\nu \text { times }} \big)
    ,
\end{equation}
where the $c_{\mu n}$'s are additional nonlinear parameters, and the $f_n$'s are Chebysev radial basis functions that smoothly go to zero at the potential cut-off. 
In \eqref{eq:mom-tens-desc}, $\mu$ and $\nu$ define the moment tensor level $\operatorname{lev}M_{\mu,\nu} = 2 + 4\mu + \nu$.
An MTP with a given level $\operatorname{lev}_{\rm MTP}$ is then constructed using all basis functions that can be obtained from all scalar contractions of the $M_{\mu, \nu}$'s that satisfy $\operatorname{lev}_{\rm MTP} \ge \sum_{i=0}^{N_{\rm m}} \operatorname{lev}M_{\mu_i,\nu_i}$, where $N_{\rm m}$ is the number of moment tensors included in the contraction.
For example, for an MTP of level 16 there are 92 basis functions, and for an MTP of level 18 there are 163 basis functions.

The parameters $\mtheta_\alpha$ and $c_{\mu n}$ are obtained by minimizing the MTP's predictions for energies, forces, and stresses, with respect to data coming from Density Functional Theory (DFT) calculations.
For further details regarding the training, we refer to Appendix \ref{sec:training}.

\subsection{Active learning}
\label{sec:AL-alg}

Suppose now that we are running a simulation with our MTP that has been trained on configurations from some training set $\scT$.
During the simulation, we come across new neighborhoods $\scN^\ast$ \emph{not present} in the training set and in order to assess whether we should better add the configuration containing those neighborhoods to our training set, we need active learning.
Active learning is a method of judging whether to add $\scN^\ast$ to the training set or not based on some scalar model uncertainty $\gamma$ \cite{settles_active_2010}.
The algorithm that computes $\gamma$ is called ``query strategy'', and there are a number of query strategies that have been successfully adapted for MLIPs.
For example, \citet{behler_representing_2014,zhang_active_2019} proposed a query strategy based on query-by-committee for neural network potentials, in which $\gamma$ is the standard deviation between different model predictions.
\citet{podryabinkin_active_2017} have proposed D-optimal design for MTPs, in which $\gamma$ has the meaning of an extrapolation of the potential.
\citet{jinnouchi_onthefly_2019,vandermause_onthefly_2020} proposed Bayesian active learning for the Gaussian process-based potentials, in which $\gamma$ is the predictive variance, which is naturally built into Gaussian process regression.

\subsubsection{D-optimal selection of training configurations}

In this work, we use the D-optimality criterion to select the training configurations, which is well-established for MTPs.
Hence, we consider $\gamma$ as the extrapolation grade per atom in the following.

In order to compute $\gamma$, assume for a moment an MTP that has $m$ coefficients and an active set with $m$ neighborhoods.
We then define the $m \times m$ Jacobian
\begin{equation}\label{eq:Jacobian}
    \uuA
    =
    \begin{pmatrix}
        \frac{\partial \clE(\clN_1; \umtheta)}{\partial \mtheta_1} & \cdots & \frac{\partial \clE(\clN_1; \umtheta)}{\partial \mtheta_m} \\
        \vdots & \ddots & \vdots \\
        \frac{\partial \clE(\clN_m; \umtheta)}{\partial \mtheta_1} & \cdots & \frac{\partial \clE(\clN_m; \umtheta)}{\partial \mtheta_m}
    \end{pmatrix}
    .
\end{equation}
The extrapolation grade $\gamma$ is then defined as the maximum change in the determinant of $\uuA$ if we would replace any $\clN_i$ with $\clN^\ast$.
Fortunately, we do not need to replace all $\clN_i$ with $\clN^\ast$ and compute all determinants individually, but can compute $\gamma$ conveniently as follows
\[
    \gamma
    =
    \underset{i}{\operatorname{max}} \vert c_i \vert,
    \qquad
    \text{with}
    \;
    \uc
    =
    \begin{pmatrix}
        \frac{\partial \clE(\clN^\ast; \umtheta)}{\partial \mtheta_1} & \cdots & \frac{\partial \clE(\clN^\ast; \umtheta))}{\partial \mtheta_m}
    \end{pmatrix}^\sT
    \uuA^{-1}
    .
\]
In general, our training set contains many more neighborhoods than coefficients, so, $\uuA$ would be overdetermined.
Therefore, we select those $m$ neighborhoods that maximize linear independence between the column vectors of $\uuA$ using the maxvol algorithm \cite{goreinov_how_2010}.

Now, in order to define whether to add a configuration $\{ \bmr_i \}^\ast$ to the training set, we compute the $\gamma$'s for all $\clN \in \{ \bmr_i \}^\ast$.
Following \cite{novikov_mlip_2021},
\[
    \begin{aligned}
        &\gamma \le 1 && \text{indicates interpolation,} \\
        1 < &\gamma \le 2 && \text{indicates accurate extrapolation,} \\
        2 < &\gamma \le 10 && \text{indicates still reliable extrapolation,} \\
        10 < &\gamma && \text{indicates risky extrapolation}
        .
    \end{aligned}
\]
Then, if the maximum $\gamma$ is higher than some threshold, we add this configuration to the training set.

\subsubsection{Active learning algorithm for large-scale simulations of fcc dislocations}
\label{sec:algo_description}

\begin{figure}[t]
    \centering
    \includegraphics[width=0.97\textwidth]{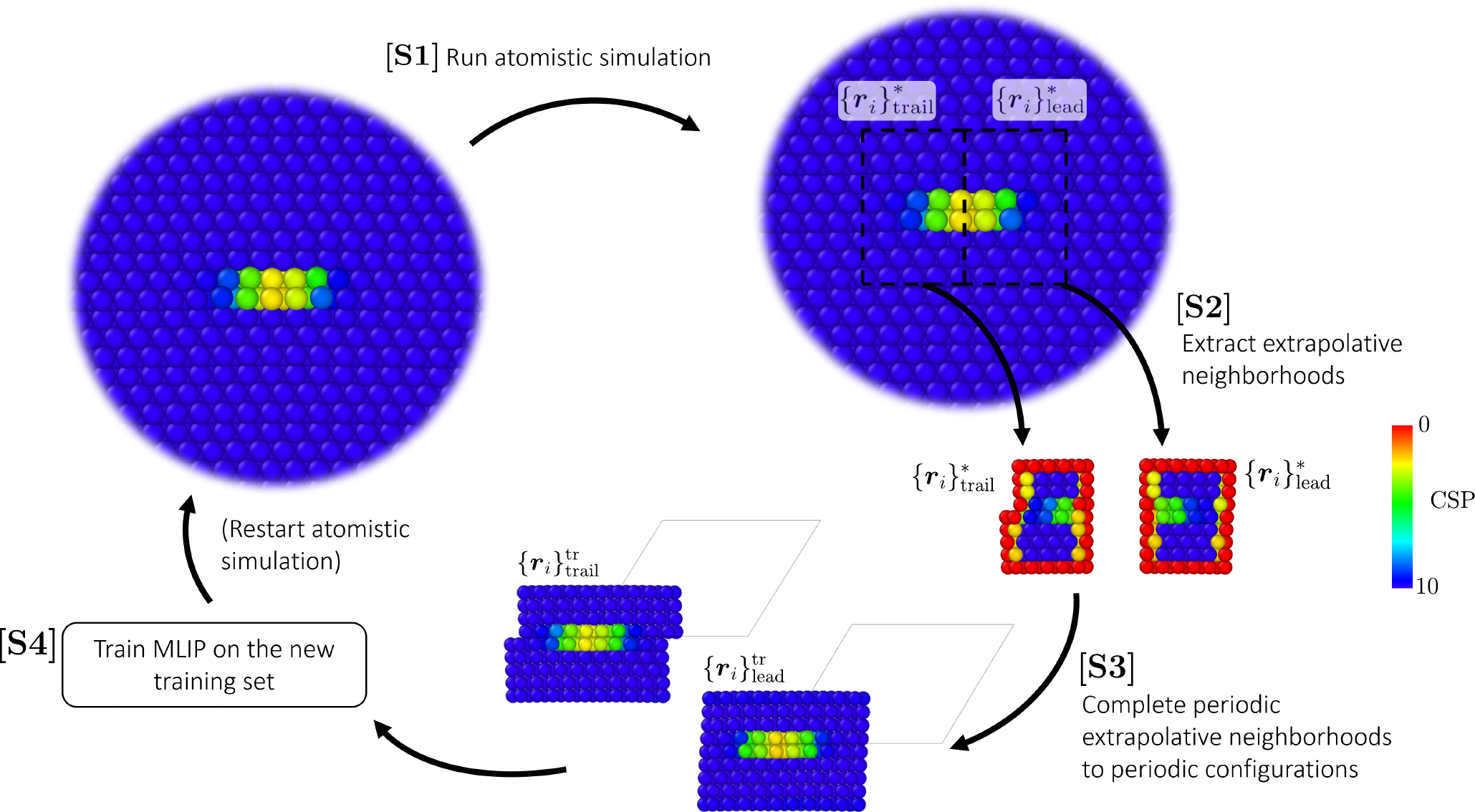}
    \caption{
    Schematic illustration of the individual steps of the active learning algorithm for fcc dislocations presented in Section \ref{sec:algo_description}.
    The coloring of the atoms is due to the centrosymmetry parameter (CSP) \cite{kelchner_dislocation_1998}, in order to visualize their deviation from the ideal crystal.}
    \label{fig:alg_scheme}
\end{figure}

In general, we are interested in simulating atomic configurations $\{ \bmr_i \}$ with tens of thousands of atoms, or even more.
While active learning can reliably detect extrapolative neighborhoods in such simulations, it is not trivial to construct training configurations containing those neighborhoods because we can not afford even one single-point plane-wave DFT calculation for such a large number of atoms.
In fact, we are only able to simulate small subsets of $\{ \bmr_i \}$ with $\sim$\,100--200 atoms at a time.
However, blindly simulating such a---\emph{generally non-periodic}---subset with plane-wave DFT degrades the accuracy of the MTP because the corresponding training configuration (that \emph{must} be subject to periodic boundary conditions) contains artificial neighborhoods at the cell boundaries (cf., \textbf{[S2]} in Figure \ref{fig:alg_scheme}).

A general way that can, in principle, be applied to any large-scale simulation, is to extract cluster configurations \cite{podryabinkin_nanohardness_2022,erhard_modelling_2023}.
However, computing clusters with plane-wave DFT requires buffer and/or vacuum regions to be introduced at the cell boundaries that can heavily increase the computational cost.
Instead of using buffer or vacuum regions, one could also optimize the atoms at the cell boundaries to mimic atomic neighborhoods close to the ones that occur in the large-scale configuration \cite{woodward_flexible_2002,hodapp_operando_2020,podryabinkin_nanohardness_2022}, but it is yet unclear how this should be done for general arrangements of atoms.

In \cite{hodapp_operando_2020}, we have developed an efficient method for screw dislocations in bcc metals.
In this method a rectangular cluster around the dislocation core is completed to a periodic training configuration by symmetrizing the atomic positions at the periodic cell boundaries.
Recently, \citet{zhang_atomistic_2022} showed that the same idea can also be applied to cracks.

In this work, we extend our algorithm for bcc screw dislocations to general dislocations in fcc metals that dissociate into two Shockley partial dislocations.
The added difficulty is that a training configuration containing a full fcc dislocation would still require too many atoms, in particular, for materials with a large separation between the two partials.
Therefore, we extend our algorithm as follows.
We first train the MTP on several bulk configurations, and configurations containing a stacking fault.
Then we are left with \emph{two regions} around each of the partial cores, $\{ \bmr_i \}^\ast_{\rm lead}$, and $\{ \bmr_i \}^\ast_{\rm trail}$, where the MTP potentially extrapolates (cf., Figure \ref{fig:alg_scheme} \textbf{[S3]}).
If the extrapolation grade in those regions exceeds some threshold, we complete this configuration to a periodic one by symmetrizing \emph{only} a rectangular configuration around each partial core---but \emph{not} around the full dislocation.
This method allows to construct two potential training configurations, $\{ \bmr_i \}^{\rm tr}_{\rm lead}$, and $\{ \bmr_i \}^{\rm tr}_{\rm trail}$, that contain the neighborhoods from the large-scale simulation region, but do not suffer from artificial neighborhoods that do not occur in the large-scale simulation region.
Details on how this procedure is implemented are postponed to Section \ref{sec:periodic_cfg}.

The full algorithm is schematically depicted in Figure \ref{fig:alg_scheme} and outlined in the following.

\begin{itemize}[leftmargin=1.8cm]
    \item[\textbf{[S0]}]
    Define the initial atomistic configuration $\{\boldsymbol{r}_i\}$ that contains a full fcc dislocation composed of two Shockley partial dislocations.
    Set the index of the first iteration to $n_{\rm start} = 1$.
    \item[\textbf{[S1]}]
    Run the simulation for $N$ iterations, starting from iteration $n_{\rm start}$.
    In each iteration $n = n_{\rm start},\ldots,n_{\rm start} + N - 1$, compute the highest extrapolation grades in both, $\{ \bmr_i \}^\ast_{\rm lead}$, and $\{ \bmr_i \}^\ast_{\rm trail}$.
    \item[\textbf{[S2]}]
    Stop the simulation after $N$ iterations---or if an extrapolation grade exceeds some threshold $\gamma_{\rm max}$.
    If any of the extrapolation grades computed in \textbf{[S1]} exceeds some threshold $\gamma_{\rm min}$, add the corresponding configuration to the \emph{set of training candidates}.
    \item[\textbf{[S3]}]
    If the set of training candidates is not empty, update the training set using the following query strategy:
    \begin{itemize}[leftmargin=1.2cm]
        \item[\textbf{[S3.1]}]
        Complete all configurations from the set of training candidates to periodic configurations using the method from Section \ref{sec:periodic_cfg}.
        \item[\textbf{[S3.2]}]
        Move the configuration with the highest extrapolation grade from the set of training candidates to the \emph{training set}.
        \item[\textbf{[S3.3]}]
        Update the Jacobian $\uuA$ \eqref{eq:Jacobian} using the maxvol algorithm.
        \item[\textbf{[S3.4]}]
        Recompute the extrapolation grades for all configurations that are left in the set of training candidates.
        If all grades are $<$\,$\gamma_{\rm min}$, go to \textbf{[S4]}, otherwise go back to \textbf{[S3.2]}.
    \end{itemize}
    \item[\textbf{[S4]}]
    If new configurations have been added to the training set, retrain the potential, and go back to \textbf{[S1]} to restart the simulation from iteration $n_{\rm start}$.
    Otherwise, set $n_{\rm start} = n_{\rm start} + N$, and go back to \textbf{[S1]} to continue the simulation.
\end{itemize}

The steps \textbf{[S1]}--\textbf{[S4]} are then repeated until convergence, or until the maximum number of iterations is reached.

\subsubsection{Completion of non-periodic extrapolative neighborhoods to periodic training configurations}
\label{sec:periodic_cfg}

Since we want to use plane-wave DFT as our ab initio model, we need to apply periodic boundary conditions on our training configurations.
However, the extrapolative configurations containing the partial dislocations, $\{\bmr_i\}^\ast_{\rm lead}$, and $\{\bmr_i\}^\ast_{\rm trail}$, shown in Figure \ref{fig:alg_scheme}, are not periodic, and computing them with plane-wave DFT would degrade the reliability of the data and the accuracy of the MTP.

\begin{figure}[hbt]
    \centering
    \includegraphics[width=0.8\textwidth]{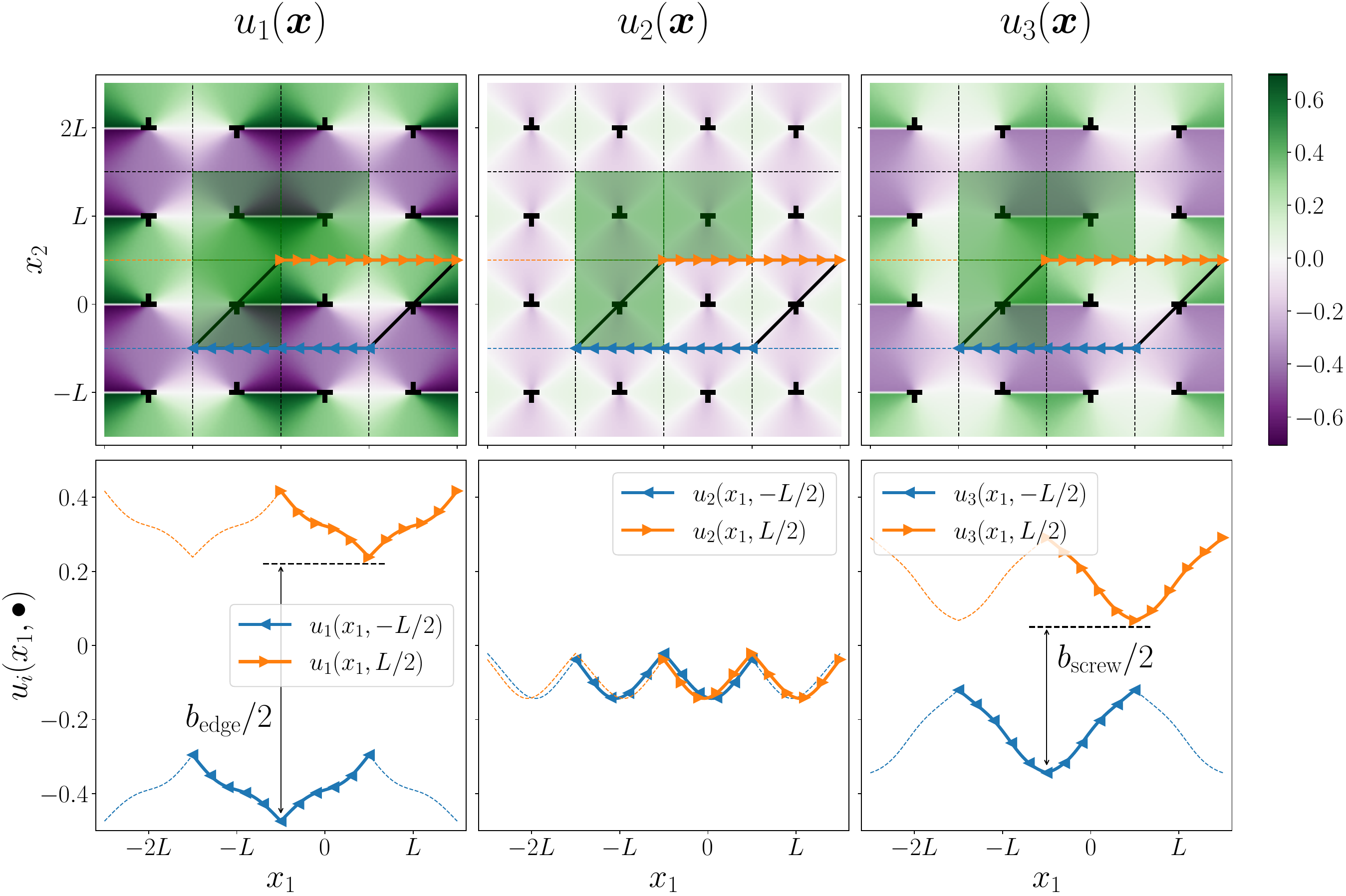}
    \caption{%
    Illustration of the solution \eqref{eq:mirrored_displ} that is periodic over a rectangular region containing four linear elastic dislocations.
    Due to symmetry, the solution is even periodic up to a constant given by half of the magnitude of the Burgers vector over the smaller triclinic cell that contains only two dislocations.}
    \label{fig:symmetrization_uel}
\end{figure}

In \citep{hodapp_operando_2020}, we have developed a method for completing such extrapolative neighborhoods to periodic configurations by symmetrizing the displacement field at the boundaries, and successfully applied it to simulate screw dislocation motion in bcc tungsten.
The algorithm works as follows.
First assume a straight dislocation along the $\rmz$-axis with the glide direction being along the $\rmx$-axis.
Let the displacement field of this dislocation be given by $\tilde\bmu(\bmx)$, with $\bmx \in \bbR^3$, such that $\tilde\bmu(\bmr_{0,i}) = \bmr_i - \bmr_{0,i}$, where $\bmr_{0,i}$ is the position of atom $i$ in its reference (bulk) configuration, with the dislocation centered at $\bmx = \bmZero$.
Next, we apply $\tilde\bmu(\bmx)$ only to the reference position of those atoms that lie in the rectangular region $[-L/2,L/2] \times [-L/2,L/2]$ with length $L$.
Now, we mirror the displacements at the boundaries of this rectangular region, and this procedure creates a displacement field in $[-3L/2,L/2] \times [-L/2,3L/2]$
\begin{equation}\label{eq:mirrored_displ}
    \bmu(\bmx)
    =
    \left\{
    \begin{aligned}
        \;
        & \tilde\bmu(x_1,x_2,x_3) && \forall\,\bmx \in [-L/2,L/2] \times [-L/2,L/2], \\
        & \tilde\bmu(-x_1 - L,x_2,x_3) && \forall\,\bmx \in [-3L/2,-L/2] \times [-L/2,L/2], \\
        & \tilde\bmu(x_1,-x_2 - L,x_3) && \forall\,\bmx \in [-L/2,L/2] \times [L/2,3L/2], \\
        & \tilde\bmu(-x_1 - L,-x_2 - L,x_3) && \forall\,\bmx \in [-3L/2,-L/2] \times [L/2,3L/2],
    \end{aligned}
    \right.
\end{equation}
which is \emph{fully periodic} (cf., Figure \ref{fig:symmetrization_uel}).

This method is general and can be applied to, in principle, any dislocation with arbitrary Burgers vector
$
    \bmb =
    \begin{pmatrix} b_{\rm edge} & 0 & b_{\rm screw} \end{pmatrix}^\sT
$
(and possibly also even to other types of defects), as shown in \ref{fig:symmetrization_uel}.
As done in \citep{hodapp_operando_2020}, we can additionally exploit the fact that the displacement field is close-to symmetric up to a constant shift $\bmb/2$ so that it suffices to consider training configurations composed of atoms located in the triclinic region
\begin{equation}\label{eq:cell_vectors_uel}
    \left\{ \bmx = \sum_i \alpha_i \bmv_i + \bmc \;\bigg\vert\; \alpha \in [0, 1] \right\},
    \qquad \text{with} \;
    \bmc =
    \begin{pmatrix} -L & 0 & -L/2 \end{pmatrix}^\sT,
\end{equation}
spanned by the cell vectors
\begin{align*}
    \bmv_1 =
    \begin{pmatrix} 2L & 0 & 0 \end{pmatrix}^\sT,
    &&
    \bmv_2 =
    \begin{pmatrix} L + b_{\rm edge}/2 & L & b_{\rm screw}/2 \end{pmatrix}^\sT,
    &&
    \bmv_3 =
    \begin{pmatrix} 0 & 0 & a_3 \end{pmatrix}^\sT,
\end{align*}
where $a_3$ is the periodic lattice spacing in the direction of the dislocation line.
In the case of elasticity, the displacement $\bmu$ is exactly periodic over the triclinic region; for a formal proof of this statement, the reader is referred to \citep{hodapp_operando_2020} (Appendix B therein).

The problem we are considering here is more complicated since there are two (partial) dislocations in our large-scale configuration.
This requires some additional steps to be performed on top of the previous method that we explain in detail below.

\begin{figure}[hbt]
    \centering
    \includegraphics[width=0.9\textwidth]{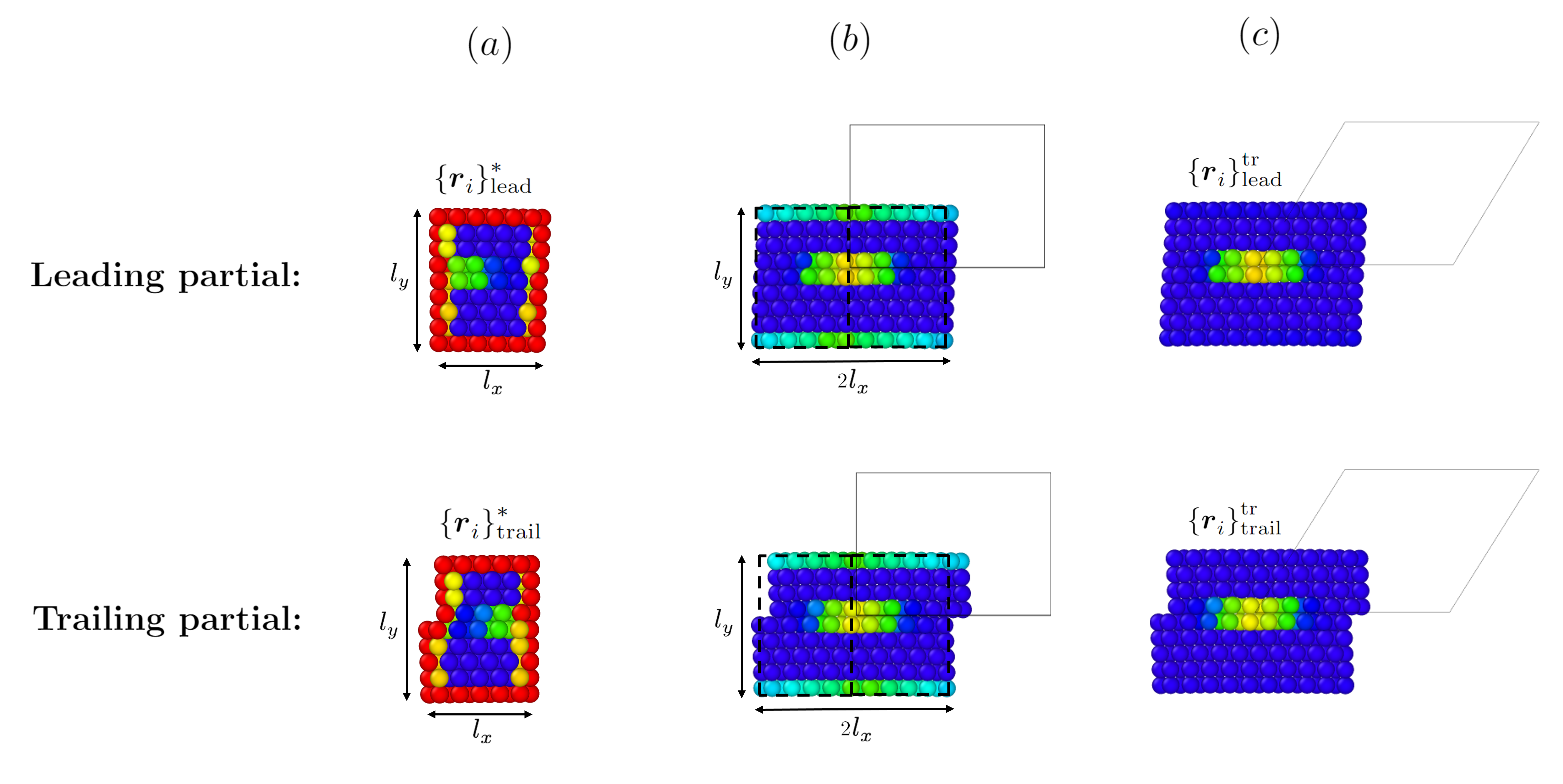}
    \caption{%
    Steps of the process of completing extrapolative neighborhoods around both partial dislocations to periodic training configurations.
    (a) Extraction of the extrapolative neighborhoods from the large-scale configuration $\{\bmr_i\}$.
    (b) Mirroring of the displacement field along the $\rmy$-axis in order to construct a configuration that is periodic along the dislocation glide direction.
    (c) Adjustment of the triclinic cell vectors to remove any artificial neighborhoods near the cell boundaries.}
    \label{fig:periodic_completion}
\end{figure}

\begin{enumerate}
    \item
    First, we detect the positions of the two partial dislocations since they may move along the glide plane during energy minimization.
    We do this by minimizing the difference between the atomistic solution and the elastic solution with the respect to the positions of the two partial dislocations.
    \item
    Next, we compute the displacement field around each of the partial cores and apply it to some rectangular subset of the reference configuration as shown in Figure \ref{fig:periodic_completion} (a).
    \item
    Now, we extend the length of this region to $2L$ and mirror the displacements along the $\rmy$-axis according to \eqref{eq:mirrored_displ}.
    This procedure creates a stacking fault in the center of the cell, but otherwise no artificial neighborhoods occur in the vicinity of the cell boundaries, except for the top and bottom layers (Figure \ref{fig:periodic_completion} (b)).
    \item
    In the final step, we modify the periodic cell vectors to remove the artificial neighborhoods at the top and bottom layers.
    Since the atomistic solution is not exactly symmetric, we compute the shift as the difference of the minima of the solutions on the top and bottom layers
    \[
        \Delta \bmu =
        \underset{x_1',z_1'}{\operatorname{min}} \, \bmu(x_1',L/2,z_1')
        -
        \underset{x_1'',z_1''}{\operatorname{min}} \, \bmu(x_1'',-L/2,z_1'')
        .
    \]
    The cell vector $\bmv_2$ is then given by
    \[
        \bmv_2 =
        \begin{pmatrix} L + \Delta u_1 + d & L & \Delta u_3 \end{pmatrix}^\sT,
    \]
    where $d$ is some additional shift that is required to match the stacking sequence and depends on the size of the cell;
    if $\bmu$ would be the elastic solution, then $\Delta \bmu = -\bmb/2$, as in \eqref{eq:cell_vectors_uel}.
    In practice, the final configuration with the modified cell vector $\bmv_2$ does not contain any artificial neighborhoods (cf., Figure \ref{fig:periodic_completion} (c)).
\end{enumerate}

\begin{figure}[t!]
    \centering
    \includegraphics[height=6cm]{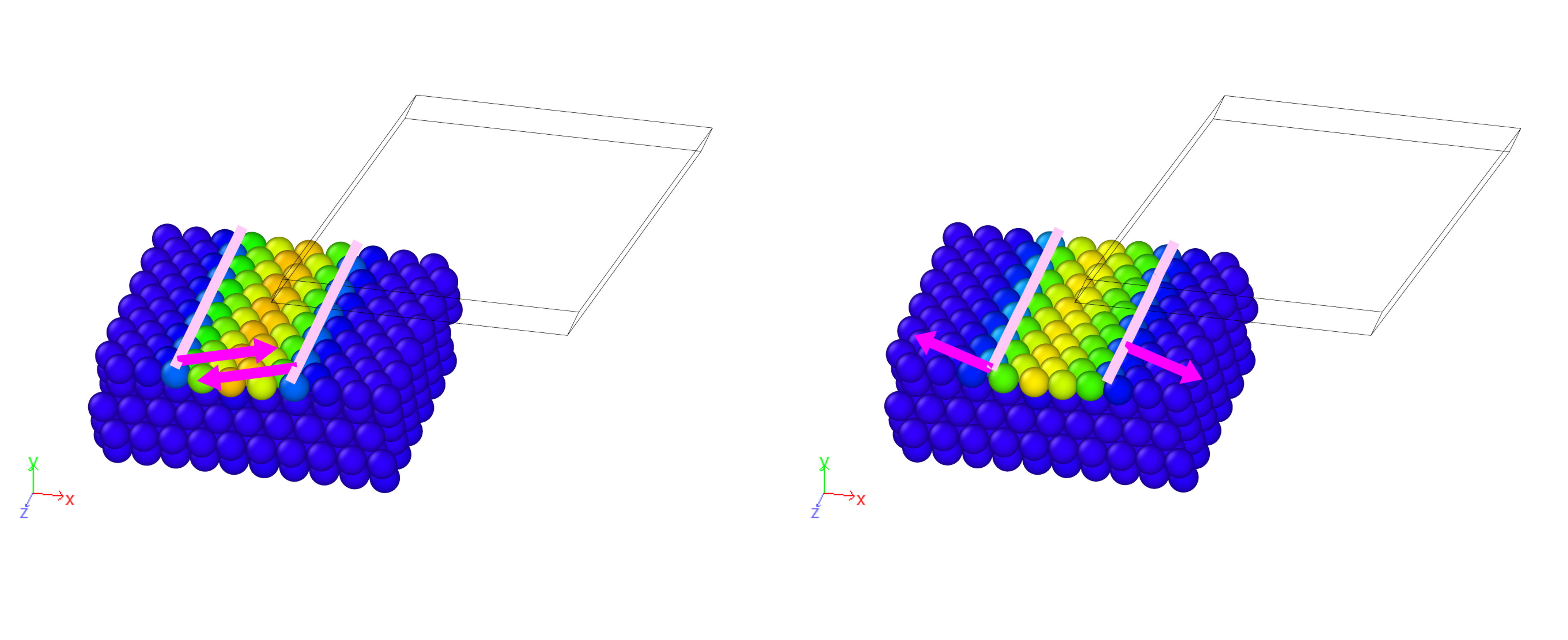}
    \caption{%
    Final training configurations $\{\bmr_i\}^\ast_{\rm lead}$ (left) and $\{\bmr_i\}^\ast_{\rm trail}$ (right) for an edge dislocation.
    The dislocations have been detected using the Dislocation Extraction Algorithm (DXA) \cite{stukowski_automated_2012}, implemented in OVITO \cite{stukowski_visualization_2010}.
    The net Burgers vector is zero over each configuration, which follows from mirroring the configuration at the $\rmy\rmz$-plane (cf., Figure \ref{fig:periodic_completion}).
    Note that, for a better visualization, the configuration has been extended in the $\rmz$-direction; the actual training configuration is confined to the triclinic cell.}
    \label{fig:training_cfgs}
\end{figure}

In each iteration, we thus create two training candidate configurations containing two dislocations that have equal and opposite partial Burgers vectors (cf., Figure \ref{fig:training_cfgs}).
Of course, the configurations themselves are not physically meaningful, however, we point out that we do \emph{not} attempt running a full simulation on them, but \emph{only} single-point DFT calculations.
Hence, if the influence of electronic interactions decays sufficiently fast, these configurations are suitable for training a local model of interatomic interaction.
We will show in the following section that the algorithm outlined above allows to construct DFT-accurate MLIPs for fcc dislocations with any character angle.

We further remark that constructing training sets this way is supported by recent analysis of \citet{ortner_framework_2023} who proved that the MLIP's error in a large-scale simulation with respect to a local DFT model converges exponentially with the size of the periodic training configurations (provided that the MLIP approximates the training set sufficiently well).
This implies that our way of training \emph{directly on neighborhoods extracted from the large-scale simulations} should lead to a nearly optimal training set.

\section{Computational results}

\subsection{Training protocol}
\label{sec:training_protocol}

We now validate the proposed algorithm for dislocations in fcc aluminum that dissociate into $1/6\langle112\rangle$ Shockley partial dislocations.
We apply the training algorithm to three types of dislocations: edge dislocations, screw dislocations, and mixed dislocations with a character angle of 30$^\circ$.
Our large-scale simulation region is a cylindrical configuration of atoms with radius 35$b$, where $b$ is the magnitude of the Burgers vector.
The simulation region contains $\sim$\,9\,500 atoms, for the edge and the mixed dislocation, and $\sim$\,5\,500 atoms for the screw dislocation.
Outside the simulation region we use Dirichlet boundary conditions, that is, we set the displacement of the atoms to the linear elastic solution of the corresponding dislocation.
As our ab inito model, we use DFT with plane-wave basis sets, as implemented in the Vienna Ab initio Simulation Package (VASP) \cite{kresse_efficient_1996}.
The corresponding simulation parameters are given in Section \ref{sec:dft_calculations}.

We perform structural relaxation using the Fast Inertial Relaxation Engine (FIRE) \cite{bitzek_structural_2006}, as implemented in the Atomic Simulation Environment (ASE) \cite{hjorthlarsen_atomic_2017}.
We consider a configuration as relaxed when the maximum absolute force on an atom is less than 1e-03\,eV/\AA.

For the training simulations, we use level-16 MTPs.
Prior to running the simulations, we train those MTPs on ten 32-atom bulk configurations and five 72-atom configurations containing a stacking fault (including random perturbations of the atoms).
We then start all our simulations with an initial partial splitting of 3.5$b$.
During the simulation, we detect the two partial dislocations, compute the extrapolation grades of the atoms in the vicinity of the partial cores, and construct the potential training configurations if one per-atom extrapolation grade exceeds a threshold of $\gamma_{\rm min} = 2$;
after $N = 30$ iterations, or if one per-atom extrapolation grade exceeds $\gamma_{\rm max} = 10$, we stop the simulation and update the training set according to step \textbf{[S3]} of the algorithm presented in Section \ref{sec:algo_description}.
The training configurations contain 180 atoms, in case of the edge and mixed dislocations, and 126 atoms, in case of the screw dislocation, respectively.

Upon convergence, the training sets of the three MTPs contained in total 24 configurations for the edge dislocation, 44 configurations for the screw dislocation, and 30 configurations for  the mixed dislocation.
Hence, training the three potentials only required 68 single-point DFT calculations in total, which underlines the efficiency of the proposed training algorithm.
Moreover, the training simulations can be run in parallel; each of them required $\sim$\,5--10 hours on a single 128-core node on the Vienna Scientific Cluster.

After training the three MTPs, we combine the training data into one big training set that now contains 68 configurations.
We then train a level-18 MTP on this combined training set and rerun the simulations with \emph{active learning switched off}.
The training errors of this MTP are shown in Table \ref{tab:training_errors_dft}.
The accuracy for per-atom energies is excellent, well below 1\,meV/atom and, therefore, close to limit of accuracy that can be achieved with interatomic potentials (cf., \cite{zuo_performance_2020}).
The force and stress errors are also within the range of high accuracy, i.e., of order of 10\,meV/{\AA} for forces, and of the order of 10e-1\,GPa for stresses.

\begin{table}[hbt]
    \centering
    \begin{tabular}{|c|x{1.5cm}|x{1.5cm}|}
        \hline
        Quantity & MAE & RMSE
        \\ \hline\hline
        Energy [eV/atom] & 4.2e-4 & 7.0e-4
        \\ \hline
        Forces [eV/\AA] & 2.9e-2 & 3.6e-2
        \\ \hline
        Stress [GPa] & 3.1e-1 & 4.8e-1
        \\ \hline
    \end{tabular}
    \caption{%
    Errors of the level-18 MTP that has been trained on configurations computed with DFT containing, i.a., edge, screw, and mixed dislocations, that were found by our active learning algorithm, as described in Section \ref{sec:training_protocol}.
    The training errors are close to the accuracy limit of interatomic potentials, indicating that the MTP can simultaneously predict the core structures of those three dislocations.}
    \label{tab:training_errors_dft}
\end{table}

In the following, we analyze the core structure and dislocation energies predicted by this level-18 MTP.
Prior to training on DFT, we have tested our algorithm by training on Embedded Atom Method (EAM) potentials.
The corresponding results are in excellent agreement with the exact solutions and can be found in Appendix \ref{sec:validation_eam}.
Considering that the DFT training errors (Table \ref{tab:training_errors_dft}) are not significantly worse than the EAM training errors (Table \ref{tab:training_errors_eam}) is one indication that our training algorithm captures the right training data, and the MTP trained on DFT data with our active learning algorithm should also give very accurate results.

\subsection{DFT calculations}
\label{sec:dft_calculations}

In the following, we compare the core structures of the edge and screw dislocation, predicted by our MTP, with the core structures predicted by the quantum mechanics/molecular mechanics (QM/MM) method of \citet{woodward_prediction_2008}.
We have used a DFT setup similar to theirs, namely, we have used projector-augmented wave pseudopotentials \cite{blochl_projector_1994,kresse_ultrasoft_1999} within the generalized gradient approximation \cite{perdew_generalized_1996}.
Moreover, we use an energy cut-off of 480\,eV, a Gaussian smearing of 0.08\,eV, and a minimum $\bmk$-point spacing of 0.15\,\AA$^{-1}$.
Electronic relaxation is performed using the preconditioned minimal residual method.
We consider a configuration as converged when the energy difference between two subsequent iterations is less than 5e-07\,eV.
With these parameters we obtain a lattice constant of 4.042\,\AA.
For setting up the boundary conditions using the linear elastic solution of a dislocation \citep{hirth_theory_1982}, we have used a finer grid of $\bmk$-points (36\texttimes36\texttimes36) to accurately predict the elastic constants, in agreement with real experiments (cf., Table \ref{tab:elastic_constants}).

In addition, we compare the dislocation core structures and dislocation energies, predicted by our MTP, to the orbital-free DFT (OF-DFT) method of \citet{iyer_electronicstructure_2015,das_electronic_2017} that uses the local density approximation for the exchange–correlation energy \citep{perdew_selfinteraction_1981}, and the Goodwin–Needs–Heine pseudopotential \citep{goodwin_pseudopotential_1990}.
This OF-DFT method uses the Wang–Govind-Carter kinetic energy functional \citep{wang_orbitalfree_1999}, which has been shown to be in good agreement with (Kohn-Sham) DFT for bulk properties and vacancy formation energies of aluminum (see, e.g., \citep{carling_orbitalfree_2003}).

\subsection{Core structure and partial splitting}

We first compare the core structures of the edge and screw dislocation, predicted by our MTP, with the core structures predicted by pure-DFT methods, namely, the quantum mechanics/molecular mechanics (QM/MM) method of \citet{woodward_prediction_2008}, and the orbital-free DFT (OF-DFT) method of \citet{iyer_electronicstructure_2015,das_electronic_2017}. For comparison, we use the Nye tensor methodology of \cite{hartley_representation_2005}, and differential displacements (cf., \cite{vitek_theory_1974}), in order to estimate the splitting distance between the partial dislocations.

\begin{figure}[t!]
    \centering
    \includegraphics[width=0.95\textwidth]{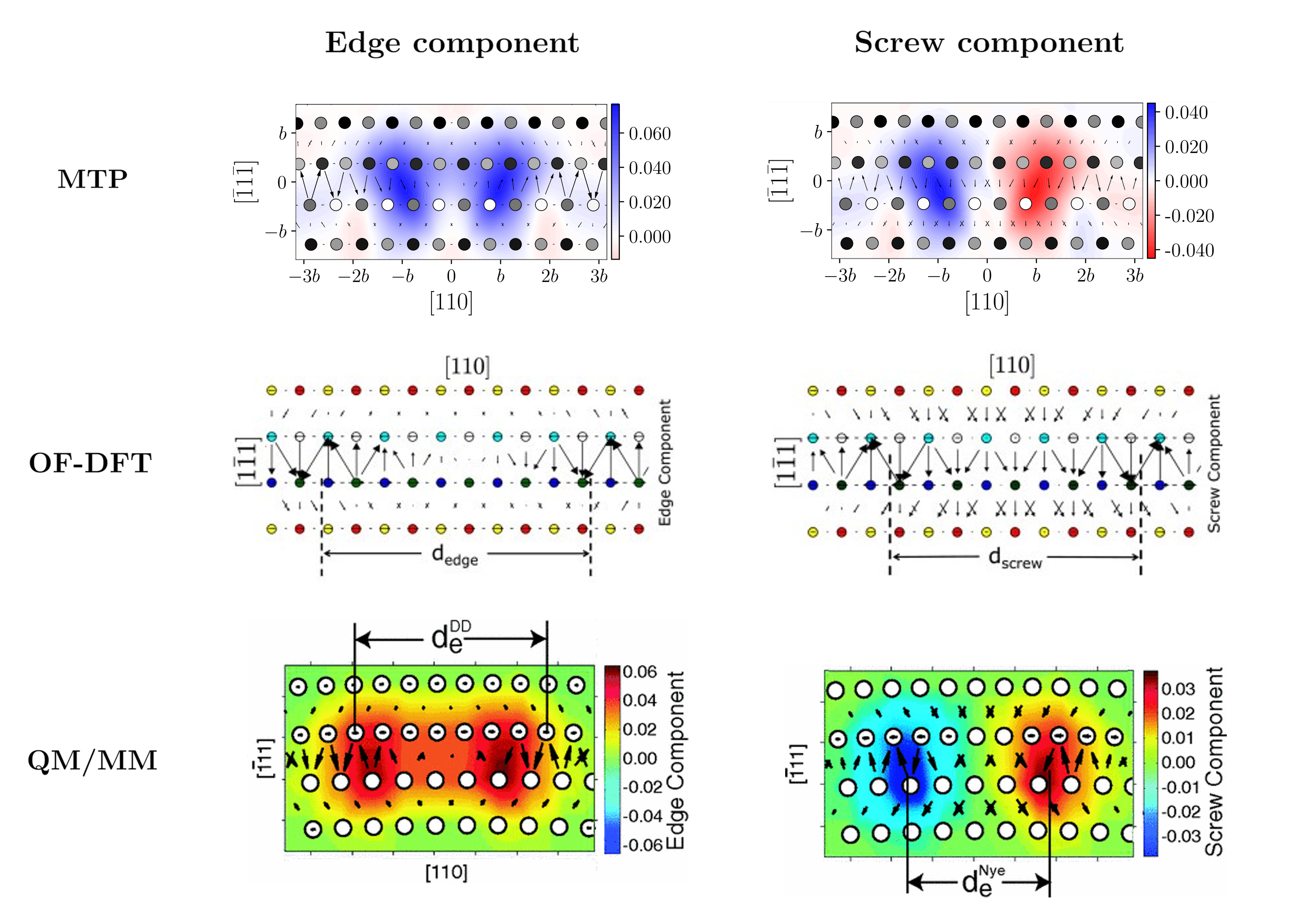}
    \caption{%
    Visualization of the edge dislocation core structure predicted by the MTP, QM/MM \cite{woodward_prediction_2008}, and OF-DFT \cite{iyer_electronicstructure_2015}, using the edge and screw components of the Nye tensor, and differential displacements.
    Overall, the agreement between the MTP and DFT core structures is very good, and the splitting distances between the partial dislocations closely coincide, up to the usual uncertainty of about the distance between two atomic planes.
    The QM/MM and OF-DFT figures are reproduced from \cite{woodward_prediction_2008} and \cite{iyer_electronicstructure_2015} with permission from APS and Elsevier.}
    \label{fig:disloc_core_edge}
\end{figure}

\begin{figure}[t!]
    \centering
    \includegraphics[width=0.95\textwidth]{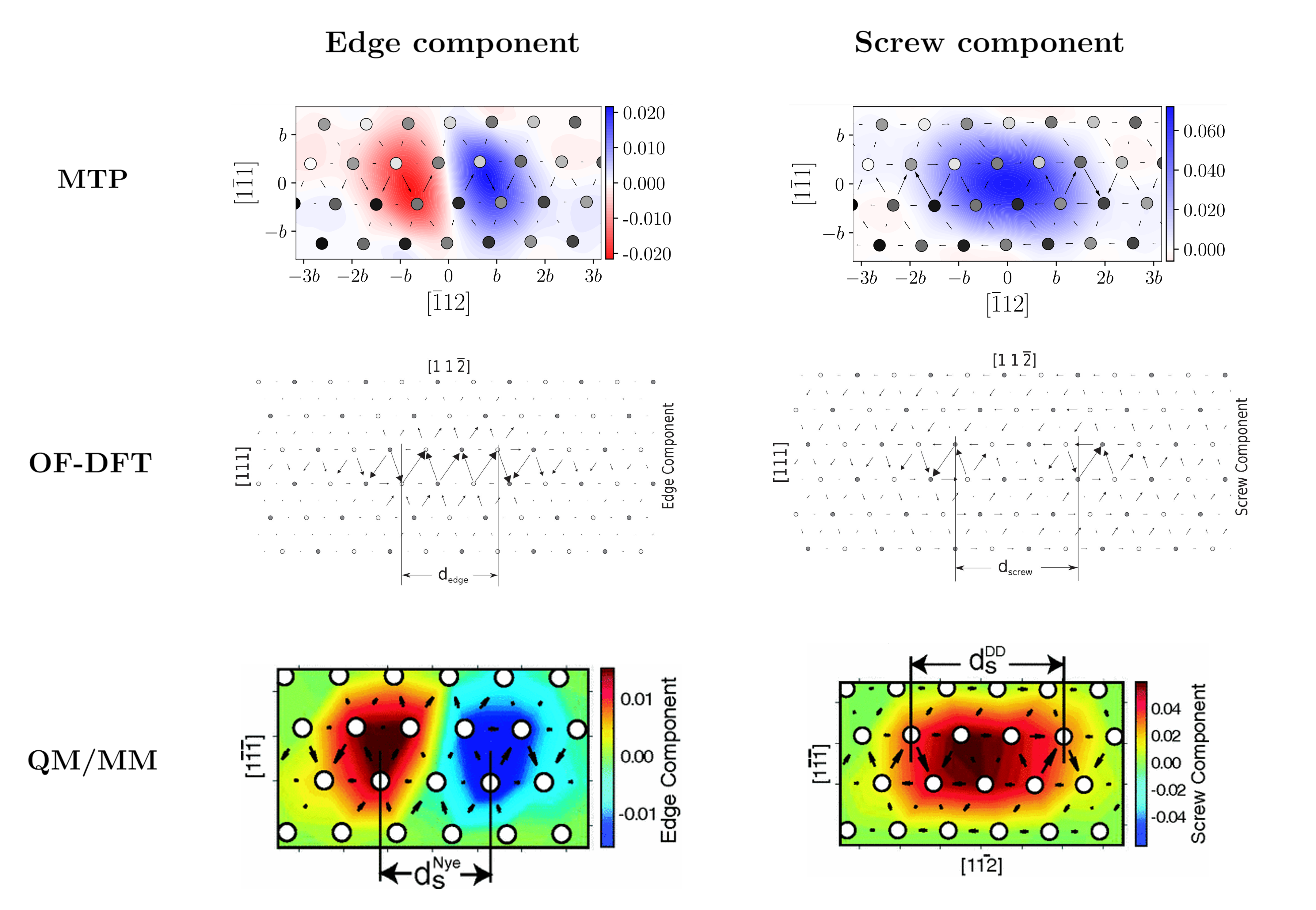}
    \caption{%
    Visualization of the screw dislocation core structure predicted by the MTP, QM/MM \cite{woodward_prediction_2008}, and OF-DFT \cite{das_electronic_2017}, using the edge and screw components of the Nye tensor, and differential displacements.
    Overall, the agreement between the MTP and DFT core structures is very good, and the splitting distances between the partial dislocations coincide remarkably.
    The QM/MM and OF-DFT figures are reproduced from \cite{woodward_prediction_2008} and \cite{das_electronic_2017} with permission from APS and Elsevier.}
    \label{fig:disloc_core_screw}
\end{figure}

From Figure \ref{fig:disloc_core_edge} and \ref{fig:disloc_core_screw}, it follows that the Nye tensor distributions are comparable for both, the edge, and the screw dislocation.
The splitting distance of the edge dislocation is computed by taking the distance between the locations of the two extrema of the screw component of the Nye tensor along the glide direction, and the splitting distance of the screw dislocation is computed by taking the distance between the locations of the two extrema of the edge component of the Nye tensor along the glide direction.
Our results for both, the edge, and the screw, coincide with those computed by \citet{woodward_prediction_2008} (cf., Table \ref{tab:partial_splitting_dft}), who used the same methodology for computing the Nye tensor.

\begin{table}[t!]
    \centering
    \begin{tabular}{|c|x{1.5cm}|x{1.5cm}|x{1.5cm}|x{1.5cm}|x{1.5cm}|x{1.5cm}|}
        \hline
        \multirow{2}{*}{Dislocation} & \multicolumn{2}{c|}{Nye tensor} & \multicolumn{3}{c|}{Differential displacements} & DXA
        \\ \cline{2-7}
        & MTP & QM/MM & MTP & QM/MM & OF-DFT & MTP
        \\ \hline\hline
        Edge & 6.6 & 7.0 \cite{woodward_prediction_2008} & 14.3 & 9.5 \cite{woodward_prediction_2008} & 12.8 \cite{iyer_electronicstructure_2015} & 15.0
        \\ \hline
        Screw & 4.6 & 5.0 \cite{woodward_prediction_2008} & 8.2 & 7.5 \cite{woodward_prediction_2008} & 8.2 \cite{das_electronic_2017} & 8.6
        \\ \hline
    \end{tabular}
    \caption{%
    Partial splitting distances (in \AA) for the edge and screw dislocation predicted by the MTP and various DFT methods; the splitting distances are computed using three different methods: the Nye tensor method, differential displacements, and the DXA.
    Overall, the agreement between MTP and DFT is very good, in particular, in view of the uncertainty of about the distance between two atomic planes spacing that is inherent in all dislocation detection methods.}
    \label{tab:partial_splitting_dft}
\end{table}

Using differential displacements, the agreement between MTP, QM/MM, and OF-DFT, is very good for the screw dislocation.
For the edge dislocation, the MTP splitting is close to the OF-DFT splitting (up to the numerical uncertainty of $\sim$\,a distance of atomic planes in the glide direction due to the discreteness of the problem), but is slightly larger than the QM/MM prediction.
While this difference is still small, well within the range of tolerable deviation, it motivates further discussion.
One possibility for the difference can be the size of the DFT region used in QM/MM.
The size of the DFT region in \cite{woodward_prediction_2008} is less than a hundred of atoms, while ours and OF-DFT contain more than a thousand of atoms.
In general, for coupled methods, such as QM/MM, there can be non-negligible spurious effects on the motion of the dislocation up to $\sim$\,3--4$b$ from the boundary \cite{dewald_analysis_2006}, and those effects may also influence the partial splitting.
\citet{olmsted_lattice_2001} have analyzed the spurious boundary stress on dislocations, and their results show that this boundary stress is much larger for edge dislocations than for screw dislocations.
This result supports our argument of attributing the differences in the partial splitting to the size of the DFT region since the splitting distances coincide when computed using the Nye tensor method, which uses the screw component of the Nye tensor, and by the fact that the splitting distance coincide across different model predictions for the screw dislocation, where the edge components of the partial dislocations are much smaller than the screw components.

Given that different results can be found in the literature, e.g., \citet{lu_electrons_2006} report a partial splitting of 5.6\,{\AA} for the edge dislocation, shows that this topic seems not to be completely settled yet.
However, we anticipate that our methodology of using MLIPs and active learning, in addition to emerging mathematical analysis \cite{ortner_framework_2023}, now provides a systematic \emph{and} tractable way for analyzing dislocation core structures in fcc metals.

We have also computed a 30$^\circ$-mixed dislocation using our MTP.
The predicted core structure and partial splitting is in agreement with the previous results for the edge and screw dislocations, i.e., the partial splitting is in between the partial splitting for the edge and screw dislocations.
The corresponding results are reported in Appendix \ref{sec:disloc_core_mixed}.

\subsection{Energy differences}

\begin{figure}[t!]
    \centering
    \begin{minipage}{0.4\textwidth}
        \centering
        \textbf{Edge}
    \end{minipage}
    \begin{minipage}{0.4\textwidth}
        \centering
        \textbf{Screw}
    \end{minipage}\\[0.8em]
    \begin{minipage}{0.4\textwidth}
        \centering
        \includegraphics[width=0.8\textwidth]{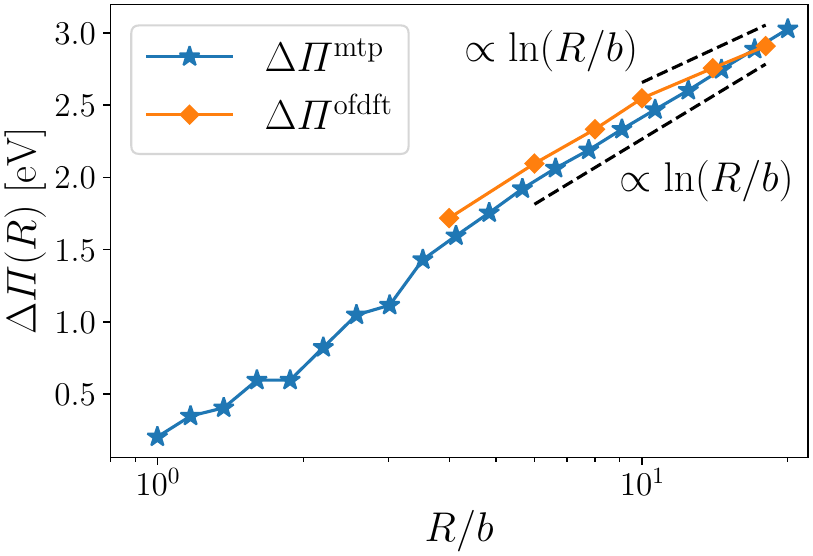}
    \end{minipage}
    \begin{minipage}{0.4\textwidth}
        \centering
        \includegraphics[width=0.8\textwidth]{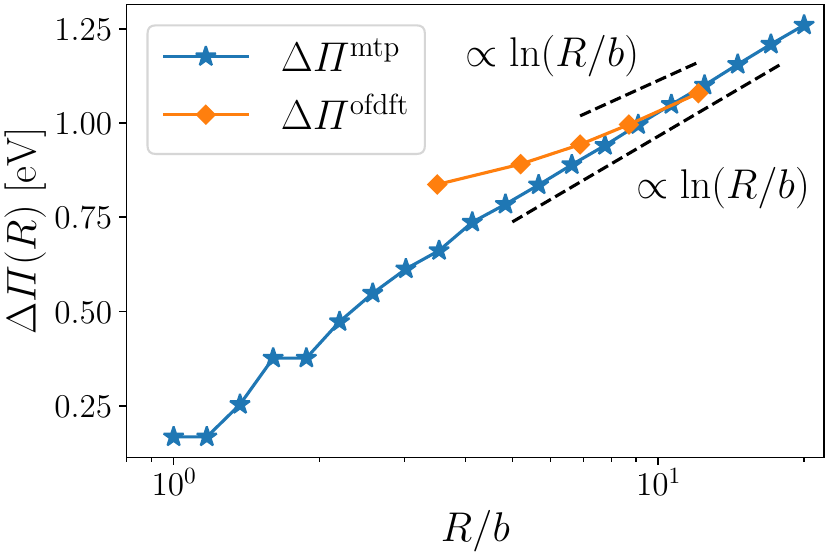}
    \end{minipage}
    \caption{%
    Dislocation energies for the edge and screw dislocation predicted by the MTP and OF-DFT.
    The agreement between the MTP and OF-DFT is very good and the far-field behavior corresponds to the dislocation energy predicted by linear elasticity (dashed lines) up to some constant core energy.}
    \label{fig:ediff_edge&screw_dft}
\end{figure}

In order to validate whether our MTP can predict energy differences, we compute the dislocation energy $\Delta\Pi = \Delta\Pi(R)$, i.e., the difference between the energy of a configuration with and without the dislocation in a cylindrical region with radius $R$ around the dislocation.
Our results for the edge and screw dislocation and those reported in \citep{iyer_electronicstructure_2015,das_electronic_2017} for OF-DFT are shown in Figure \ref{fig:ediff_edge&screw_dft}.

For the edge dislocation, the agreement between the MTP and OF-DFT dislocation energies is very good.
The small difference between 4\,$\le$\,$R/b$\,$\le$\,10 could be due to the boundary conditions used for OF-DFT that are set up with the isotropic elastic constants, which differ from the anisotropic ones (cf., Table \ref{tab:elastic_constants}).
This is supported by the fact that the OF-DFT curve is approximately linear in $\operatorname{ln}(R/b)$ between 4\,$\le$\,$R/b$\,$\le$\,10 and its the slope, i.e., the energy factor \citep{hirth_theory_1982}, is very close to the one of the MTP curve in this interval.
Another reason could be due electronic effects present up to $R/b$\,$\approx$\,10, as argued in \citep{iyer_electronicstructure_2015}, that can not be captured with our local MTP.

For the screw dislocation, the MTP and OF-DFT results differ qualitatively for small $R/b$ between 3\,$\le$\,$R/b$\,$\ge$\,7.
Again, this could be due to the isotropic boundary conditions used for OF-DFT.
Another reason could be the smaller simulation regions used for OF-DFT;
in \citep{iyer_electronicstructure_2015,das_electronic_2017} the simulation region was gradually increased with increasing $R$, so $\Delta\Pi^{\rm ofdft}$ corresponds to the total energy of a configuration with radius $R$, whereas we use a simulation region of 35$b$ and sum up the per-atom energies up to some $R$.
As shown in \citep{hodapp_coupled_2018}, the difference between isotropic and anisotropic boundary conditions can have a nonnegligible influence on dislocations close to the boundary, even for materials like aluminum that are only weakly anisotropic.
Moreover, there can be electronic effects up to 7\,$R/b$, as pointed out in \citep{das_electronic_2017}, that can not be reproduced with a local MLIP.
The behavior for smaller $R/b$ could be similar to results from \citep{dan_firstprinciples_2022} who reported oscillating dislocation energies close to the dislocation core for screw dislocations in magnesium using a QM/MM method.

Most importantly, in both cases, the dislocation energies for OF-DFT and MTP agree well when the OF-DFT curve enters the linear elastic regime.
The asymptotic scalings predicted by the MTP and OF-DFT differ slightly since the isotropic elastic constants used to set up the boundary conditions for OF-DFT deviate $\sim$\,10--30\,\% from experiments (cf., Table \ref{tab:elastic_constants}).
On the other hand, the elastic constants predicted by the MTP are in excellent agreement with DFT and the experimental ones.
This implies that the MTP can predict reliable core energies.

\begin{table}[hbt]
    \centering
    \begin{tabular}{|c|x{2.8cm}|x{2.8cm}|x{2.8cm}|x{2.8cm}|}
        \hline
        Component & DFT & MTP & OF-DFT \cite{iyer_electronicstructure_2015,das_electronic_2017} & Experimental \cite{vallin_elastic_1964}
        \\ \hline\hline
        $C_{11}$ & 113.4 & 112.0 & 95.3 & 116.3
        \\ \hline
        $C_{12}$ & 60.0 & 63.4 & 51.3 & 64.8
        \\ \hline
        $C_{44}$ & 32.9 & 30.4 & 22.0 & 30.9
        \\ \hline
    \end{tabular}
    \caption{%
    Cubic elastic constants (in GPa) predicted by DFT, the MTP, OF-DFT (isotropic Voigt average), and real experiments.}
    \label{tab:elastic_constants}
\end{table}

We have also computed the dislocation energy for the 30$^\circ$-mixed dislocation.
As for the partial splitting, the results are agreement with the results for the edge and screw in the sense that the dislocation energy for the mixed dislocation lies in between those two limiting cases. 
In the future, it would be interesting to also compare those results with the dislocation energies predicted by QM/MM methods (e.g., \cite{dan_firstprinciples_2022}) for further validation.

\section{Concluding Remarks}

\subsection{Discussion and potential applications}

Constructing a MLIP which is general and predictive is commonly referred to the problem of transferability.
One successful methodology to solve the problem of transferability is active learning, but it is impractical for large-scale problems because, in this case, one cannot afford even one single-point plane-wave DFT calculation, in general.
One important example for such large-scale problems are dislocations in fcc materials.
The particular difficulty that arises in fcc materials is that the dislocations typically split into two Shockley partial dislocations, and, depending on the material, their separation can be so large that simulating the full dislocation with plane-wave DFT alone would be infeasible;
for example, \citet{deng_largescale_2023} used configurations of 7\,200 atoms for simulating screw dislocations in copper with interatomic potentials in a quadrupole arrangement.

To overcome this limitation, we have developed an active learning algorithm for training MTPs during large-scale simulations of dislocations in fcc materials.
Our active learning algorithm only extracts a small cluster of atoms around each partial core, computes the per-atom extrapolation grade of each atom, and, if one of the grades is larger than some threshold, completes the cluster to a periodic configuration of 100--200 atoms that can be conveniently computed with plane-wave DFT.
We have validated our algorithm by simulating dislocations in fcc aluminum and the MTP, trained with our algorithm, was able reproduce all existing DFT results available in the literature for the core structure, the partial splitting, and the dislocation core energy.

Hence, we anticipate that our algorithm can be readily used to construct reliable MLIPs for dislocations to be used for applications that can go beyond the scope of validity of existing empirical interatomic potentials, for example:
\begin{itemize}
    \item
    One possible application of our algorithm are dislocations in alloys.
    This, of course, may require larger training configurations than used here to accommodate solutes and/or impurities like hydrogen along the dislocation line, but still appears feasible.
    For example, our algorithm could be used to construct MLIPs for computing dislocation-solute interactions that are inputs to solute-strengthening models (cf., \cite{leyson_quantitative_2010}).
    This MLIP can then be used within workflows for predictive high-throughput screening (e.g., \citep{moitzi_initio_2023}) to find the strongest alloy over a wide range of compositions.
    \item
    Our algorithm can be readily used to construct training sets for MLIPs to be used in large-scale simulations of curved dislocations.
    This is supported by recent work of \citet{wang_theoretical_2023} who showed for bcc metals that a MLIP trained on a few configurations that contain only straight dislocations with different character angles can be used to predict energies of dislocation loops.
    \item
    Moreover, we anticipate that the proposed algorithm is not limited to fcc, but also to dissociated dislocations in other materials, such as, e.g., basal dislocations in hexagonal closed-packed magnesium.
    \item
    Another potential application where we envision our algorithm to become valuable are magnetic materials.
    Including spins into the functional form of interatomic potentials doubles the size of a neighborhood and this significantly increases the complexity of constructing a good training set because the configurational space is now much larger---but active learning can come to the rescue!
    Recently, we approached the problem of training magnetic MTPs using a DFT method in which spins can be constrained in the same way as atomic positions \citep{kotykhov_constrained_2023}.
    Within this approach, the active learning algorithm developed here can potentially be applied verbatim to dislocations in magnetic materials.
\end{itemize}

\subsection{How to simulate multi-defect interactions?}

Finally, we remark that, presently, our algorithm cannot be readily applied to problems for which defects cannot be sufficiently isolated, such as dislocations cutting through precipitates.
For such problems, we currently recommend to first train a potential on configurations that appear to be relevant for the physical problem to be solved, e.g., bulk configurations, stacking faults, anti-phase boundaries, etc., or on an existing dataset, e.g., from \href{https://www.materialscloud.org/home}{materialscloud.org}.
Afterwards, we recommend to run several simulations of straight dislocations and add additional configurations with high uncertainties to the training set using our algorithm.
When starting the actual simulation of, say, a dislocation interacting with precipitate, one may leave active learning switched-on and, if the uncertainty becomes too high in parts of the simulation region, run plane-wave DFT calculations on those parts using clusters \citep{podryabinkin_nanohardness_2022}, or periodic completion of fragments \citep{jalolov_mechanical_2023}, and add them to the training set.
Those methods can be more expensive than ours and pose the danger of introducing artificial neighborhoods polluting the training set, but if the initial training set is already diverse enough such calculations would be triggered only rarely.
In the future, another approach could be training on many small systems combined with training on a few large systems using more efficient ab initio methods like those DFT methods that use finite element basis sets to improve the MLIP's predictions for large arrangements of defects.

\section{Acknowledgments}

We thank the two unknown reviewers for their constructive comments that helped us to improve the present paper.
Moreover, we thank Franco Moitzi, Oleg Peil, Daniel Scheiber, and Andrei Ruban, for fruitful discussions.
In particular, we thank Franco for providing his code for creating the Nye tensor plots.

Financial support under the scope of the COMET program within the K2 Center “Integrated Computational Material, Process and Product Engineering (IC-MPPE)” (Project No 886385), is highly acknowledged. This program is supported by the Austrian Federal Ministries for Climate Action, Environment, Energy, Mobility, Innovation and Technology (BMK) and for Labour and Economy (BMAW), represented by the Austrian Research Promotion Agency (FFG), and the federal states of Styria, Upper Austria and Tyrol.

\section*{Appendix}

\begin{appendices}

\counterwithin*{equation}{section}
\renewcommand\theequation{\thesection\arabic{equation}}

\section{MTP training}
\label{sec:training}

Suppose we are given a training set $\scT = \{ \{ \bmr_i \}_j, \Pi_j^{\rm qm}, \{ \bmf_i \}_j^{\rm qm}, \bmsigma_j^{\rm qm} \}_{j = 1, \ldots, M}$ that contains $M$ atomic configurations $\{ \bmr_i \}_j$ and its associated quantum-mechanical energies $\Pi_j^{\rm qm}$, forces $\{ \bmf_i \}_j^{\rm qm}$,  and stresses $\bmsigma_j^{\rm qm}$.
We then identify the MTP coefficients $\scC = \{ \mtheta_\alpha, c_{\mu n} \}$ by minimizing the loss functional
\begin{equation}\label{eq:loss}
    \scL(\scT,\scC)
    =
    \sum_{j = 1}^M
    \left(
    w_\rme
    \left( \Pi_j^{\rm mtp}(\scC) - \Pi_j^{\rm qm} \right)^2
    +
    w_\rmf
    \left(
    \sum_{i = 1}^N
    \left\| \bmf_i^{\rm mtp}(\scC) - \bmf_i^{\rm qm} \right\|^2
    \right)
    +
    w_\sigma
    \left\| \bmsigma_j^{\rm mtp}(\scC) - \bmsigma_j^{\rm qm} \right\|^2
    \right)
\end{equation}
with respect to $\scC$, with the energies, forces, and stresses, being weighted as follows
\begin{align*}
    w_\rme = 1, && w_\rmf = 1\text{e-}02, && w_\sigma = 1\text{e-}03.
\end{align*}
To minimize \eqref{eq:loss}, we use SciPy's BFGS solver.
For our initial training we use a limit of 500 iterations, for retraining during active learning we use a limit of 200 iterations, and for retraining the potential on the entire training set after gathering all data we use a limit of 2\,000 iterations (cf., Section \ref{sec:training_protocol}).

\section{MTP predictions for the mixed dislocation}
\label{sec:disloc_core_mixed}

\begin{figure}[hbt]
    \centering
    \begin{minipage}{0.5\textwidth}
        \centering
        \textbf{Edge component}
    \end{minipage}\hfill
    \begin{minipage}{0.5\textwidth}
        \centering
        \textbf{Screw component}
    \end{minipage}\\[0.8em]
    \begin{minipage}{0.5\textwidth}
        \centering
        \includegraphics[width=0.8\textwidth]{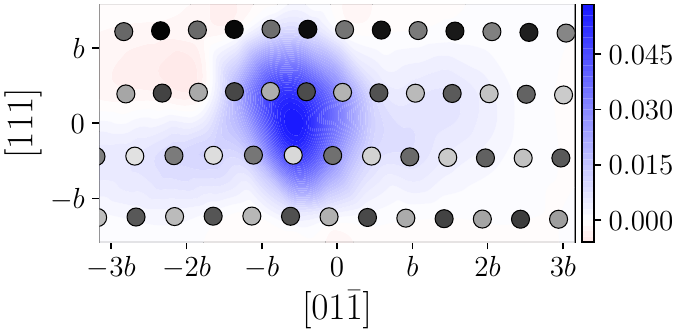}
    \end{minipage}\hfill
    \begin{minipage}{0.5\textwidth}
        \centering
        \includegraphics[width=0.8\textwidth]{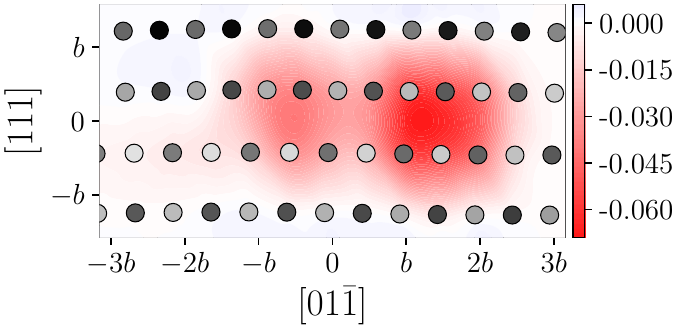}
    \end{minipage}
    \caption{%
    Visualization of the 30$^\circ$-mixed dislocation core structure predicted by the MTP using the edge and screw components of the Nye tensor.}
    \label{fig:disloc_core_mixed}
\end{figure}

The core structure of the 30$^\circ$-mixed dislocation predicted by the MTP that has been trained according to Section \ref{sec:training_protocol} is shown in Figure \ref{fig:disloc_core_mixed}.
We have computed the splitting distance of the mixed dislocation by taking the distance between the locations of the two maxima of the screw component of the Nye tensor along the glide direction.
Using this method, the splitting distance is 5.4\,\AA, which is in between the values for the edge (6.6\,\AA) and the screw (4.6\,\AA).
Using the DXA, the partial splitting of the mixed dislocation is 10.8\,\AA, which is also in between the values for the edge (15\,\AA) and the screw (8.6\,\AA).

\begin{figure}[hbt]
    \centering
    \begin{minipage}{0.333\textwidth}
        \centering
        \textbf{Mixed (30$^\circ$)}
    \end{minipage}\\[0.8em]
    \begin{minipage}{0.333\textwidth}
        \centering
        \includegraphics[width=0.97\textwidth]{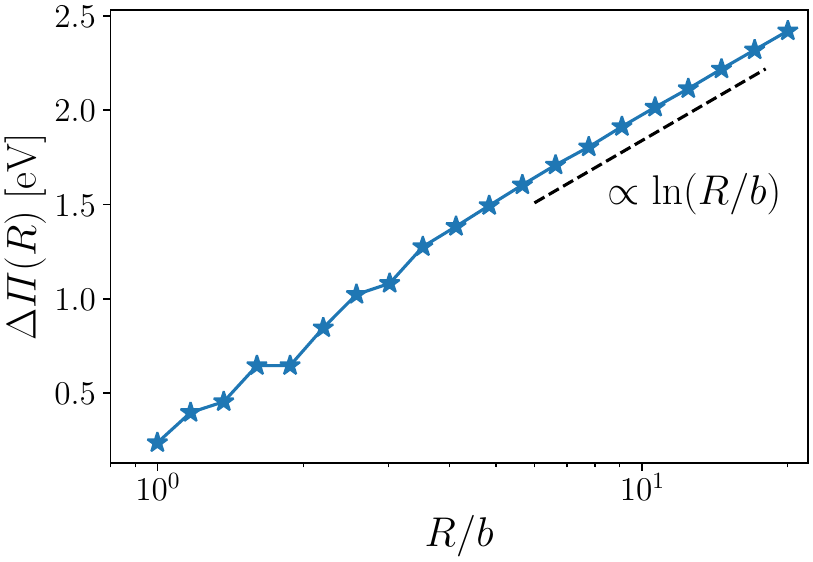}
    \end{minipage}
    \caption{%
    Dislocation energies for the mixed dislocation predicted by the MTP.
    The far-field behavior corresponds to the dislocation energy predicted by linear elasticity (dashed lines) up to some constant core energy.}
    \label{fig:ediff_mixed_dft}
\end{figure}

The dislocation energy for the mixed dislocation is shown in Figure \ref{fig:ediff_mixed_dft}.
As for the partial splitting, the result is in agreement with the dislocation energies for the edge and screw dislocations (cf., Figure \ref{fig:ediff_edge&screw_dft}).

\section{Validation of the training algorithm using EAM as a reference model}
\label{sec:validation_eam}

For validation, we also ran our training algorithm using the EAM potential of \citet{ercolessi_interatomic_1994} as a reference model instead of expensive DFT.
Following our training protocol from Section \ref{sec:training_protocol}, we first trained three separate level-16 MTPs on the edge, screw, and mixed dislocation, respectively.
Upon convergence, the training sets of the three MTPs contained 22 configurations for the edge dislocation, 33 configurations for the screw dislocation, and 28 configurations for the mixed dislocation.
After training the three MTPs, we combine the training data into one big training set that now contains 53 configurations.
We then train a level-16 MTP on this combined training set and rerun the simulations with active learning switched off.

The training errors, shown in Table \ref{tab:training_errors_eam}, are very low, which is not surprising since an MTP is able to approximate an EAM potential exactly.

\begin{figure}[hbt]
    \centering
    \includegraphics[width=0.7\textwidth]{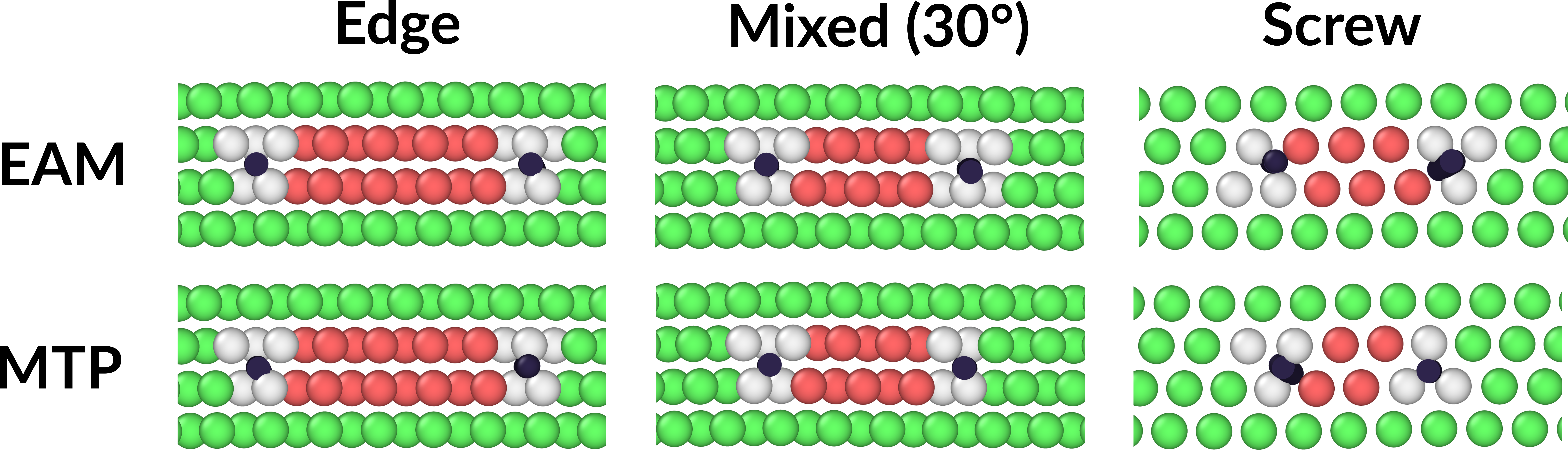}
    \caption{Visualization of the dislocation cores predicted by EAM and MTP using the common neighbor analysis and the DXA.}
    \label{fig:cna_eam_vs_mtp}
\end{figure}

The core structures predicted by the MTP and the EAM potential are visualized in Figure \ref{fig:cna_eam_vs_mtp} using the common neighbor analysis (CNA) \cite{honeycutt_molecular_1987}.
For the edge and mixed dislocation, there is no visible difference between MTP and EAM and the partial splitting distances coincide (cf., Table \ref{tab:partial_splitting_eam}).
For the screw dislocation, the MTP core appears to be slightly more narrow than the EAM core, but note that CNA assignment of structure types is very sensitive to small atomic displacements; so, both cores can be considered to be in good agreement.

\begin{figure}[H]
    \centering
    \begin{minipage}{0.333\textwidth}
        \centering
        \textbf{Edge}
    \end{minipage}\hfill
    \begin{minipage}{0.333\textwidth}
        \centering
        \textbf{Mixed (30$^\circ$)}
    \end{minipage}\hfill
    \begin{minipage}{0.333\textwidth}
        \centering
        \textbf{Screw}
    \end{minipage}\\[0.8em]
    \begin{minipage}{0.333\textwidth}
        \centering
        \includegraphics[width=0.97\textwidth]{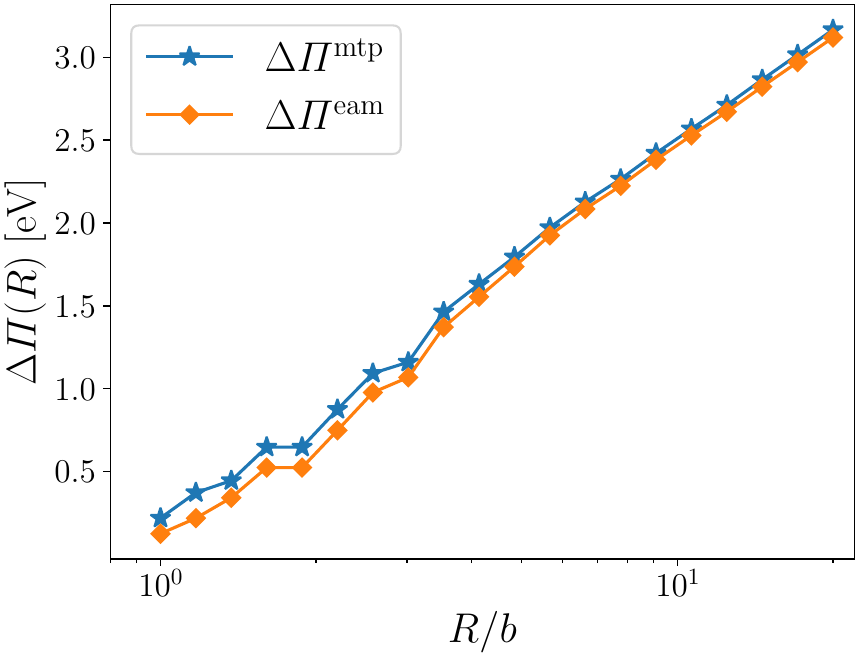}
    \end{minipage}\hfill
    \begin{minipage}{0.333\textwidth}
        \centering
        \includegraphics[width=0.97\textwidth]{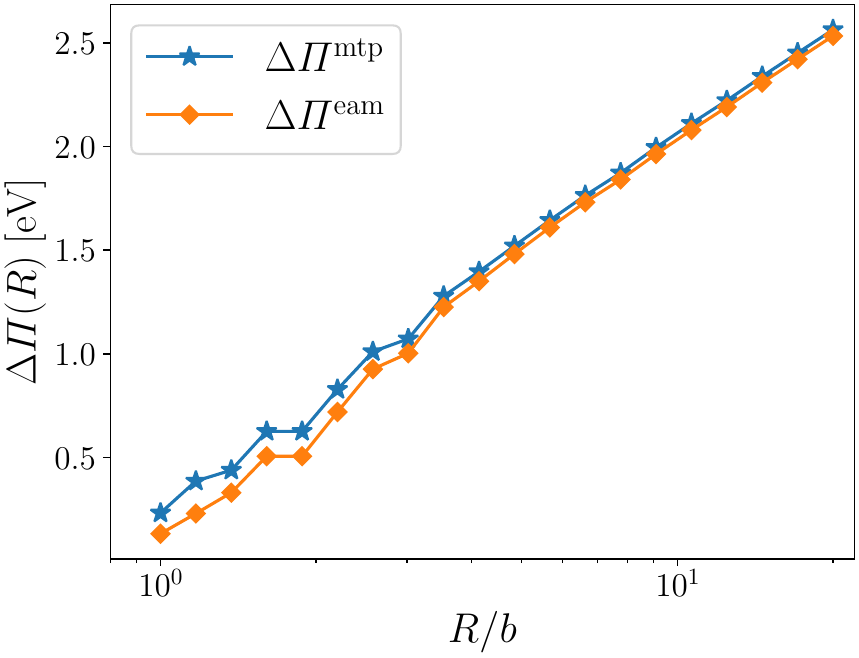}
    \end{minipage}\hfill
    \begin{minipage}{0.333\textwidth}
        \centering
        \includegraphics[width=0.97\textwidth]{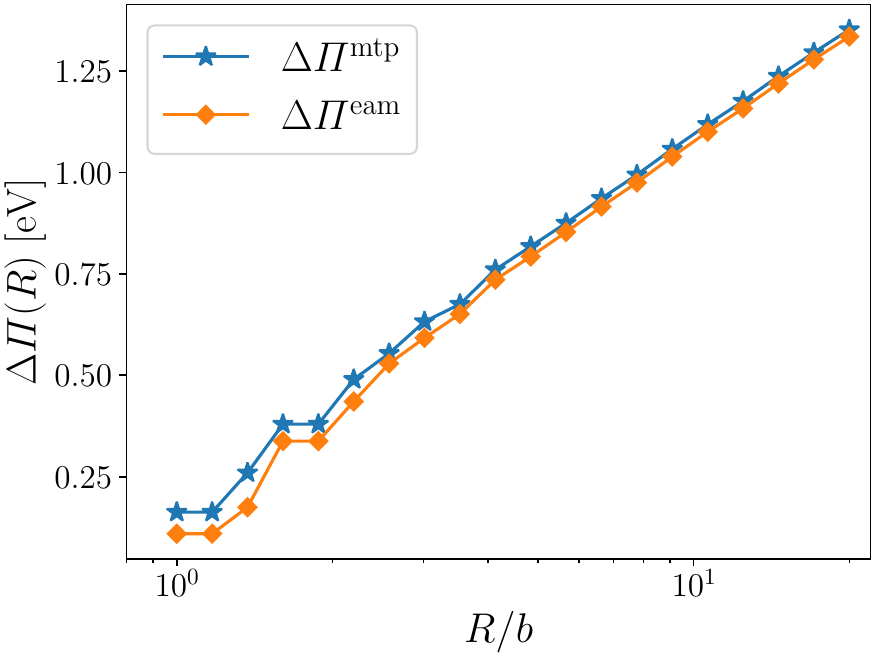}
    \end{minipage}
    \caption{%
    Dislocation energies for the edge, mixed, and screw, dislocation predicted by the MTP and the EAM potential.}
    \label{fig:ediff_eam}
\end{figure}

The corresponding dislocation energies are shown in Figure \ref{fig:ediff_eam}.
Again, the quantitative agreement between MTP and EAM is very good in the vicinity of the dislocation core;
in the far-field, they even agree almost exactly.
Moreover, the far-field behavior corresponds to the dislocation energy predicted by linear elasticity (dashed lines) up to some constant core energy.

\begin{table}[hbt]
    \centering
    \begin{tabular}{|c|x{1.5cm}|x{1.5cm}|}
        \hline
        Quantity & MAE & RMSE
        \\ \hline\hline
        Energy [eV/atom] & 1.0e-4 & 1.9e-4
        \\ \hline
        Forces [eV/\AA] & 3.5e-3 & 5.9e-3
        \\ \hline
        Stress [GPa] & 2.0e-2 & 2.7e-2
        \\ \hline
    \end{tabular}
    \caption{Errors of the level-16 MTP that has been trained on configurations computed with EAM containing, i.e., edge, screw, and mixed dislocations, that were found by our active learning algorithm, as described in Section \ref{sec:validation_eam}.}
    \label{tab:training_errors_eam}
\end{table}

\begin{table}[hbt]
    \centering
    \begin{tabular}{|c|x{1.5cm}|x{1.5cm}|}
        \hline
        Dislocation & EAM & MTP
        \\ \hline\hline
        Edge & 15.2 & 14.7
        \\ \hline
        Mixed (30$^\circ$) & 11.1 & 10.7
        \\ \hline
        Screw & 8.9 & 8.1
        \\ \hline
    \end{tabular}
    \caption{%
    DXA partial splitting distances (in \AA) for the EAM reference model and a level-16 MTP.}
    \label{tab:partial_splitting_eam}
\end{table}

\end{appendices}

\printbibliography[heading=bibintoc]

\end{document}